\documentclass[twocolumn,prl,floatfix,longbibliography,aps]{revtex4-1}
\usepackage{amsmath}
\usepackage{amssymb}
\usepackage{graphicx}
\usepackage{bm}
\usepackage{color}
\usepackage{hyperref}
\usepackage{upgreek}
\usepackage{scalerel}
\usepackage{pgffor}
\usepackage{pdfpages}
\usepackage{float}
\usepackage{lscape}   
\makeatletter
\AtBeginDocument{\let\LS@rot\@undefined}
\makeatother

\AtBeginDocument{%
	\newwrite\bibnotes
	\def\bibnotesext{Notes.bib}
	\immediate\openout\bibnotes=\jobname\bibnotesext
	\immediate\write\bibnotes{@CONTROL{REVTEX41Control}}
	\immediate\write\bibnotes{@CONTROL{%
			apsrev41Control,author="08",editor="1",pages="1",title="0",year="0"}}
	\if@filesw
	\immediate\write\@auxout{\string\citation{apsrev41Control}}%
	\fi
}%

\begin{document}

\title{Supercurrent detection of topologically trivial zero-energy states in nanowire junctions}
\author{Oladunjoye A. Awoga}
 \author{Jorge Cayao} 
 \email[ ]{jorge.cayao@physics.uu.se}
 \author{Annica M. Black-Schaffer}
\affiliation{Department of Physics and Astronomy, Uppsala University, Box 516, S-751 20 Uppsala, Sweden}

\date{\today}
\begin{abstract}
 We report the emergence of zero-energy states in the trivial phase of a short nanowire junction with strong spin-orbit coupling and magnetic field, formed by strong coupling between the nanowire and two superconductors.
The zero-energy states appear in the junction when the superconductors induce a large energy shift in the nanowire, such that the junction naturally forms a quantum dot, a process that is highly tunable by the superconductor width. Most importantly, we demonstrate that the zero-energy states produce a $\pi$-shift in the phase-biased supercurrent, which can be used as a simple tool for their unambiguous detection, ruling out any Majorana-like interpretation.
\end{abstract}\maketitle

Majorana bound states (MBSs) in topological superconductors  have generated remarkable interest  due to their 
potential applications in fault tolerant quantum computation \cite{kitaev2001unpaired,RevModPhys.80.1083,PhysRevLett.121.267002}.
A promising route for engineering the topological phase is based on nanowires (NWs) with strong Rashba spin-orbit coupling (SOC) and proximity-induced $s$-wave superconductivity, with MBSs emerging at the NW ends for sufficiently large magnetic fields \cite{PhysRevLett.105.077001,PhysRevLett.105.177002,Alicea:PRB10}.
Initial issues, such as a soft superconducting gap \cite{Lee:PRL12,Pientka:PRL12,Bagrets:PRL12,Liu:PRL12,Rainis:PRB13,Sau:13,Zitkoetal},
of the first experiments \cite{Mourik:S12, xu, Das:NP12,Finck:PRL13, Churchill:PRB13, lee2014spin} have been solved through the fabrication of high quality interfaces between the NW and external superconductors (SCs)~\cite{chang15,Higginbotham,Krogstrup15,zhang16,Albrecht16,Deng16,Nichele17,Suominen17,Chen2017experimental,Marcus2018nonlocality,zhang18}.

Despite the advances, there is still no consensus whether MBSs have been observed or not. In fact, recent reports show that trivial zero-energy Andreev bound states (ABSs) from e.g.~chemical potential inhomogeneities, appearing well outside the topological phase
\cite{Prada:PRB12,Cayao2015SNS,JorgeEPs,Fer18}, can also lead to a $2e^{2}/h$ quantized conductance \cite{StickDas17,vuik2018reproducing}, a feature previously attributed solely to MBSs \cite{Law:PRL09}. This controversy can at least partially be attributed to oversimplified models used to describe the experiments. Indeed, a common treatment of superconductivity has been to simply add an induced superconducting gap into a one-dimensional (1D) NW model, ignoring all other effects caused by coupling a SC to a NW.

A more accurate approach is to study the whole NW+SC system, since the achieved high-quality interfaces result in a strong coupling between NW and SC and thus the SC generates both an induced gap and affect other NW parameters.
Importantly, the NW energies are shifted when the coupling between the SC and NW is strong due to the lowest states having a large weight in the SC~\cite{Stanescu2017Proximity,PhysRevB.96.125426,PhysRevB.97.165425, ReegBeilstein, PhysRevB.98.245407}. This results in an effective chemical potential $\mu_{\rm eff}$ in the NW, which regulates when the NW reaches the topological phase. Therefore using a NW+SC model is crucial for gaining further insights into the experimental situation.

In this Letter we study the whole NW+SC system and find trivial zero-energy ABSs spontaneously emerging in a NW strongly coupled to two SCs forming a short superconductor-normal-superconductor (SNS) junction. The zero-energy ABSs appear in the junction when the SCs induce a large $\mu_{\rm eff}$ in the NW, such that the junction forms natural quantum dot (QD).  The QD formation occurs at regular intervals, every Fermi wavelength increment in SC width, and is thus predictable. By simply regulating the width of the SCs, we can tune the NW from an ideal regime with no energy shifts, to forming a QD or even a potential barrier (PB) at the junction. The formation of the QD and its zero-energy ABSs is therefore very different from previous situations where the QD was simply put in by hand~\cite{Liu2011Detecting,droste2012josephson,Vernek2014Subtle,Ruiz2015Interaction,Ptok2017Controlling, Liu2017QDot, vuik2018reproducing}.
Most importantly, we find that the trivial zero-energy QD states produce a $\pi$-shift in the phase-biased supercurrent, while MBSs appearing in the topological phase do not. Thus the Josephson effect in short SNS junctions offers a remarkably powerful, yet simple tool for distinguishing between trivial zero-energy states and MBSs.

\begin{figure}[!t]
	\centering
	\includegraphics[width=.48\textwidth]{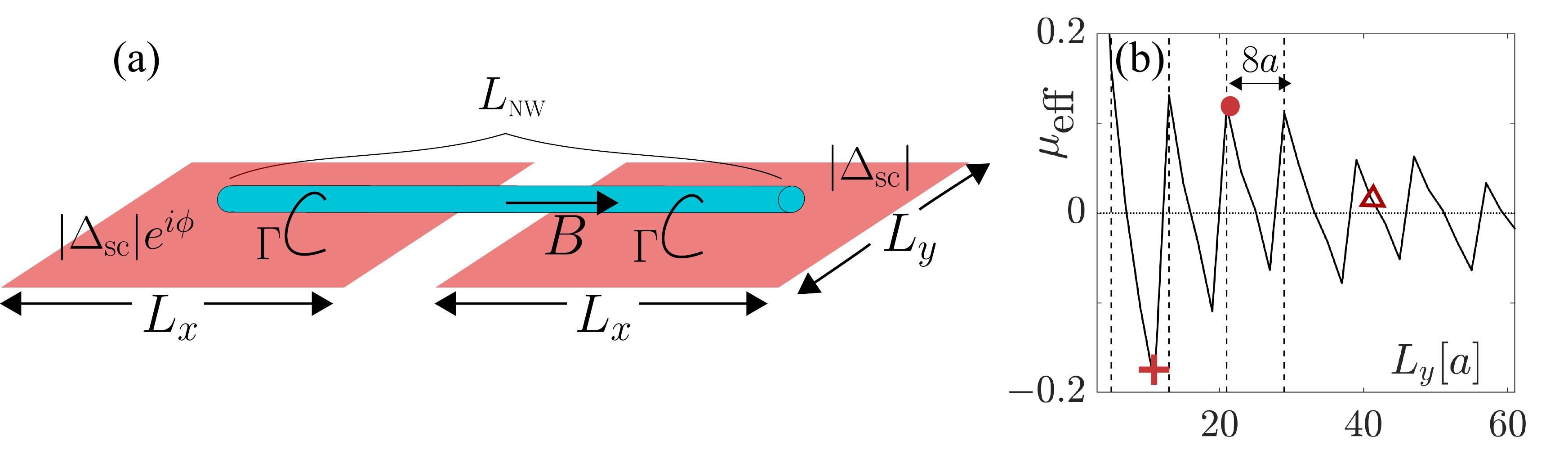}
	\caption{(a) 1D NW (cyan) coupled to the middle of two 2D SCs (red) by $\Gamma$. A short central region of the NW is left uncoupled, giving a short SNS junction with a $\phi$  superconducting phase difference. (b) Effective chemical potential profile deep into the S parts of the NW as a function of SC width $L_y$. Markers are representative points at which three cases are studied: ideal (triangle), PB (cross), and QD (dot).}
	\label{Fig1}
\end{figure}

\emph{Model}.--- We use a 1D NW with strong SOC with the right (R) and left (L) parts strongly coupled to  the middle of two 2D conventional SCs, leaving only the central part of the NW uncoupled and forming a short SNS junction, see Fig.\,\ref{Fig1}(a). By varying a magnetic field parallel to the NW we easily tune the topology of the junction.
The Hamiltonian is thus $\mathcal{H}= \mathcal{H}_{\scaleto{\rm NW}{3.5pt}} + 
\mathcal{H}_{\scaleto{\rm SC}{3.5pt}}^{\scaleto{\rm L}{4.5pt}}+\mathcal{H}_{\scaleto{\rm SC}{3.5pt}}^{\scaleto{\rm R}{4.5pt}} +\mathcal{H}_{\scaleto{\rm S-W}{3.5pt}}$, with
\begin{equation}
\begin{split}
\mathcal{H}_{\scaleto{\rm NW}{3.pt}}&=\sum_{x=1,\sigma\sigma'}^{L_{\scaleto{\rm NW}{3.pt}}} d^{\dagger}_{x\sigma}\left( \varepsilon_{\scaleto{\rm NW}{3.pt}}\delta_{\sigma\sigma'}  + B\sigma_{\sigma\sigma'}^x\right)d_{x\sigma'}
-  \sum_{x=1,\sigma}^{L_{\scaleto{\rm NW}{3.pt}}-1} \\
& d^{\dagger}_{x\sigma} \left( \right.\left. t_{\scaleto{\rm NW}{3.pt}} \delta_{\sigma\sigma'} - i\alpha_{\scaleto{\rm NW}{3.pt}}\sigma_{\sigma\sigma'}^y\right) d_{x + 1,\sigma'} + \text{H.c.} \,,\\
\mathcal{H}_{\scaleto{\rm SC}{3.5pt}}^{\scaleto{\rm R/L}{6.pt}}&=\sum_{\mathrm{i},\mathrm{j},\sigma}^{}c^{\dagger}_{\mathrm{i}\sigma}\left[\left( \varepsilon_{\rm sc} \delta_{\mathrm{i}\mathrm{j}}-t_{\rm sc}\delta_{\langle \mathrm{i},\mathrm{j} \rangle}\right) c_{\mathrm{j}\sigma}
+\Delta_{\rm sc}^{\scaleto{\rm R/L}{6.pt}}(\mathrm{i})c^{\dagger}_{\mathrm{i}\uparrow}c^{\dagger}_{\mathrm{i} \downarrow}\right]  + {\rm H.c.} ,\,\\
\mathcal{H}_{\scaleto{\rm S-W}{3.5pt}}&=-\Gamma\sum_{\rm i}\sum_{x=1,\sigma}^{L_{\scaleto{\rm NW}{3.pt}}}c^{\dagger}_{{\rm i}\sigma}d_{x\sigma}\delta_{{i_y},\frac{L_y+1}{2}}\delta_{{i_x},x}+{\rm H.c.} \nonumber,
\end{split}
\end{equation}
where $d_{{x}\sigma}$ is the destruction operator for a particle with spin $\sigma$ at site $x$   in the $L_{\scaleto{\rm NW}{3.5pt}}$ long NW, while $c_{\mathrm{i}\sigma}$  is the destruction operator at site $\mathrm{i}=(i_x,i_y)$  in the 2D SCs with length $L_x$, width $L_y$. Here, $\langle \cdots \rangle$ implies nearest neighbor sites, $t$ represents the nearest neighbor hopping and $\mu$ the chemical potential, such that the on-site energies $\varepsilon_{\scaleto{\rm NW}{3.pt}}=2t_{\scaleto{\rm NW}{3.pt}}-\mu_{\scaleto{\rm NW}{3.pt}}$, $\varepsilon_{\rm sc}=4t_{\rm sc}-\mu_{\rm sc}$. In the NW $\alpha_{\scaleto{\rm NW}{3.pt}}=\alpha_{\rm R}/2a$ is the SOC, with $\alpha_{R}$ the SOC strength and $a$ the lattice constant, and $B$ is the effective Zeeman coupling caused by the magnetic field with $\sigma^\nu$ a Pauli matrix. The SCs have an onsite $s$-wave superconducting order parameter $\Delta_{\rm sc}^{\scaleto{\rm R/L}{6.pt}}(\mathrm{i})=|\Delta_{\rm sc}|{\rm e}^{i\phi_{\scaleto{\rm R/L}{6.pt}}}$, with  $\phi_{\scaleto{\rm R/L}{6.pt}}$ being the SC phase. Finally, $\mathcal{H}_{\scaleto{\rm S-W}{3.5pt}}$ is the NW-SC tunneling Hamiltonian with finite coupling strength $\Gamma$, whenever the NW touches either SCs.  
%

We solve the Hamiltonian within the Bogoliubov-de Gennes framework~\cite{DeGennes} using parameters in units of  $t_\textrm{sc}$: $\mu_\textrm{sc}=0.5$, $\mu_{\scaleto{\rm NW}{3.pt}}=0.02t_{\scaleto{\rm NW}{3.pt}}$, $\alpha_{\scaleto{\rm NW}{3.pt}}=0.05t_{\scaleto{\rm NW}{3.pt}}$, $t_{\scaleto{\rm NW}{3.pt}}=4$, which accounts for the small NW effective mass and mismatching Fermi wavevectors in NW and SC, and being close to realistic values. We also set $\Delta_\textrm{sc}(\mathrm{i})=0.1$, $\phi_{\scaleto{\rm R}{5.pt}}=0$, and $\phi_{\scaleto{\rm L}{5.pt}}=\phi$. Here, the strong coupling regime, with the induced gap in the NW close to $\Delta_{\rm sc}$, is reached around  $\Gamma=0.7$. For smaller $\Delta_\textrm{sc}$ and  $\mu_{\rm sc}$, a smaller $\Gamma$ achieves strong coupling. 
Further, we use $L_x=520a$, $L_{\scaleto{\rm NW}{3.pt}}=1000a$, and keep the N-junction  $2a$ long, to reach realistic sizes 
with the outer ends of the NWs well within the SCs. 
The width of the SC, $L_y$, is varied in order to tune the influence of the SC on the NW~\cite{,PhysRevB.96.125426,PhysRevB.97.165425, ReegBeilstein, PhysRevB.98.245407}.
We have verified that our results remain qualitatively unchanged for $\Delta_\textrm{sc}$ and $\Gamma$ both being smaller (or even larger), as well as when $\Delta_\textrm{sc}(\mathrm{i})$ is calculated self-consistently~\cite{BlackSchaffer2008, bjornson2016piphase, awoga2017disorder, theiler2018majorana, Mashkoori2019Majorana}. Our results also do not depend on $L_x$, $L_{\scaleto{\rm NW}{3.pt}}$, junction length, provided $L_x\,, L_{\scaleto{\rm NW}{3.pt}}$ are longer than the superconducting coherence length and the junction is short, see Supplementary Material (SM) for more information \cite{SM}. 

As a result of strong coupling to the SC, all  inherent NW parameters are renormalized \cite{Stanescu2017Proximity,PhysRevB.96.125426,PhysRevB.97.165425, ReegBeilstein, PhysRevB.98.245407,woods2019electronic}. 
Most important is an energy shift of the NW bands~\cite{PhysRevB.98.245407}. We encode this by an effective chemical potential $\mu_{\rm eff}$, which we define as the energy of the bottom of the hybridized subband closest to the Fermi energy (since superconductivity occurs around the Fermi energy). We extract $\mu_{\rm eff}$ deep in the S regions of the NW and find that it oscillates as a function of $L_y$, see Fig.\,\ref{Fig1}(b). The oscillations are due to a mismatch between the SC and NW bands, with period (here $8a$) given by the SC  Fermi wavelength. Thus, by changing $L_y$ we can easily tune through a range of $\mu_{\rm eff}$.

\begin{figure}[!t]
	\centering
	\includegraphics[width=0.48\textwidth]{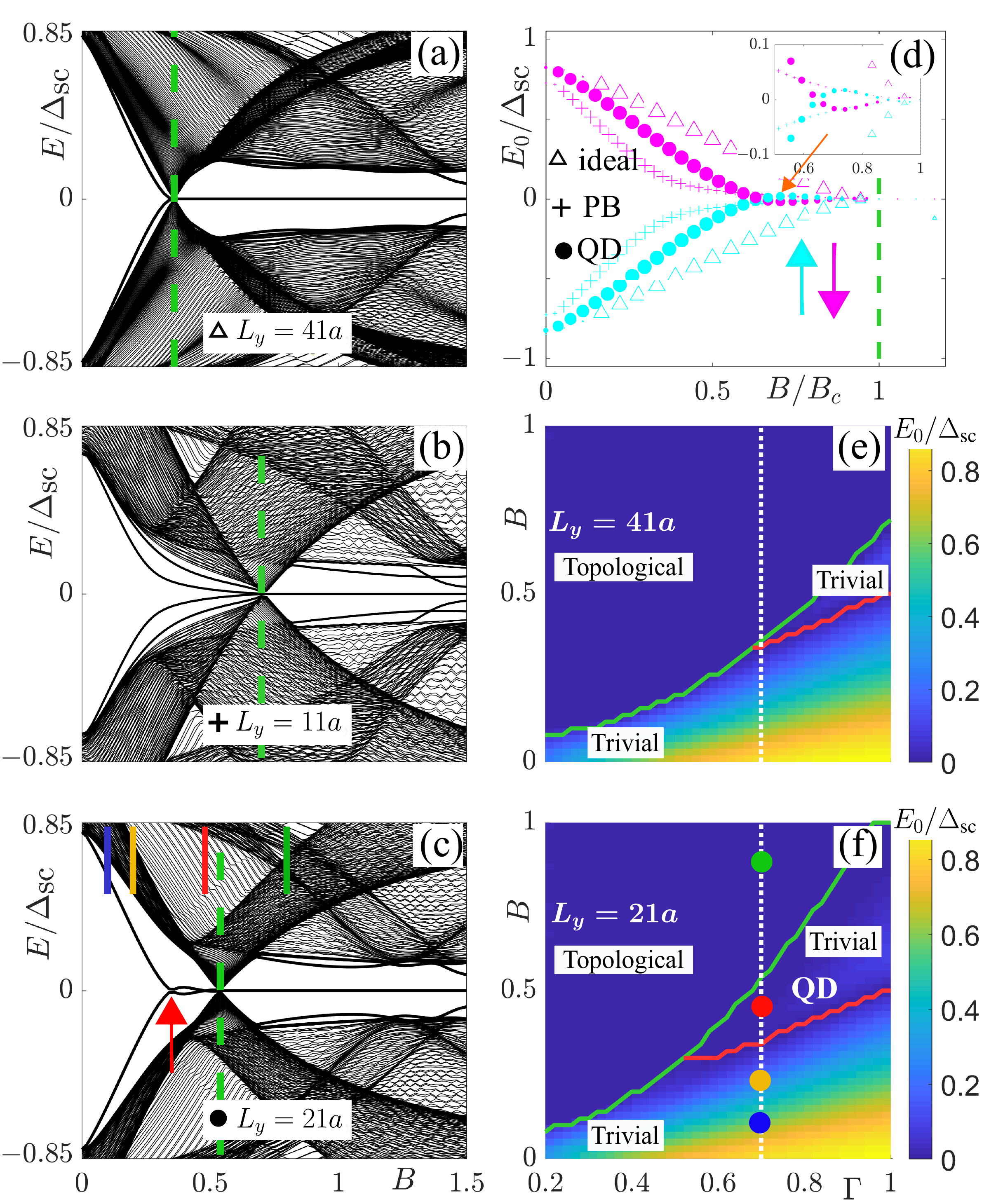}
	\caption{Zeeman field-dependent spectrum at $\phi=0$ for ideal ($L_y = 41a$) (a), PB ($L_y =11a$) (b) and QD ($L_y = 21a$) (c) cases. Vertical dashed green lines in (a-d) mark topological phase transition, while red arrow in (c) marks start of zero-energy levels.  (d) Local spin projection at the junction $S_{x={L}_{\scaleto{\rm NW}{3.pt}}/2}^{(x)}$  in lowest level $E_0$ for cases (a-c) as a function of   $B/B_{c}$. Cyan/magenta marks spin up/down while marker size denotes magnitude. (e,f) Color plot of $E_0$ as a function of $\Gamma$ and $B$ for $L_y=41a$ (e) and $L_y=21a$ (f) cases. Green line marks topological phase transition, red line start of the supercurrent $\pi$-shift, and dotted white line $\Gamma =0.7$. Filled circles in (f) denotes colored markings in (c).}
	\label{Fig2}
\end{figure}
%
\emph{Low-energy spectrum}.--- 
When the S regions of the NW get a non-zero $\mu_{\rm eff}$, the properties of the SNS junction change. 
We show this first by studying the Zeeman dependent low-energy spectrum  at $\phi=0$ for three values of the SC width $L_y$, see Fig.\,\ref{Fig2}(a-c).
The common characteristic in all three cases is that the spectrum exhibits a sizable gap at zero $B$, indicating the presence of superconductivity, which then closes and reopens at the critical field $B_{\rm c}$ signaling the topological phase transition (green dashed line).  By calculating the topological invariant for a NW coupled to a single SC \cite{Alexandradinata2014Wilson} we verify that the gap closure in Fig.\,\ref{Fig2}(a-c) matches the topological phase transition point. In the topological phase the SNS system hosts a pair of  MBSs, with zero energy,  one at each end of the NW (outer MBS), for all cases.
Since $\mu_\textrm{eff}$ changes the NW properties, we find that $B_{\rm c}$ also changes somewhat with $L_y$.

Remarkably, there is a very strong effect of $L_{y}$ on the low-energy spectrum inside the junction, resulting in the emergence of additional low-energy states below $B_{\rm c}$. These can be understood when comparing $\mu_{\rm eff}$ in the S regions of the NW to the native chemical potential $\mu_{\scaleto{\rm NW}{3.pt}}$, which is still the relevant energy in the N region. 
In fact, in Fig.\,\ref{Fig2}(a) the low-energy spectrum does not exhibit any unusual features, since here $\mu_{\rm eff}\approx \mu_{\scaleto{\rm NW}{3.pt}}$ (triangle in Fig.\,\ref{Fig1}(b)). We refer to this regime as the ideal case. 
However, when $\mu_\textrm{\rm eff}< \mu_{\scaleto{\rm NW}{3.pt}}$ (cross in Fig.~\ref{Fig1}(b)), the junction acts as a potential barrier (PB) and we see in Fig.~\ref{Fig2}(b) that such PB junction can host discrete low-energy  levels in the trivial phase.
Finally, when $\mu_\textrm{\rm eff} > \mu_{\scaleto{\rm NW}{3.pt}}$ (dot in  Fig.~\ref{Fig1}(b)), there is instead a quantum dot (QD) profile in the junction. Remarkably, this QD accommodates a clear single zero-energy crossing in the trivial phase, see Fig.~\ref{Fig2}(c). 

We here stress that the QD with a zero-energy crossing in the trivial phase emerges spontaneously at the junction, just due to strong NW-SC coupling and tuning $L_y$. 
We have numerically verified that the QD zero-energy states occur for $\tilde{\alpha} <B <B_c$, where $\tilde{\alpha}$ is the renormalized SOC in the NW (dependent on $L_y$ and $\Gamma$, here $\tilde{\alpha}\approx 0.5 \alpha_{\scaleto{\rm NW}{3.pt}}$), see SM \cite{SM}. 
Zero-energy states have previously been reported in simple 1D models with a QD put in by hand~\cite{droste2012josephson,Lee2013Kondo,Cayao2015SNS, Ptok2017Controlling, Liu2017QDot,vuik2018reproducing}, producing signatures similar to MBSs and thus challenging attempts trying to distinguish between such trivial zero-energy levels and MBSs~\cite{PhysRevB.98.245407, vuik2018reproducing,schrade2018andreev,yavilberg2019differentiating}.
In our work the QD instead develops naturally and we also find that the trivial zero-energy crossings appear solely in the QD regime, not in the PB or ideal regimes.

Further insights can be obtained from the local spin projection along $B$ (i.e. the $x$-component), in the lowest level $E_0$ states, which is given by $S_{x}^{(x)} = v_{x\uparrow}^*u_{x\downarrow} +  u_{x\downarrow}^*v_{x\uparrow}$, and superscript/subscript denotes component/position and $u_{x\sigma},v_{x\sigma}$ are the wave function amplitudes at position $x$ \cite{Sticlet2012Spin,Bjorson2014spin,Szumniak2017Spin,Serina2018boundary}.  In Fig.~\ref{Fig2}(d) we show $S_{x}^{(x)}$ at the junction, i.e.~$x = L_{\scaleto{\rm NW}{3.pt}}/2$, with marker size denoting the magnitude. $S_{L_{\scaleto{\rm NW}{3.pt}}/2}^{(x)}$ vanishes in the topological phase as the lowest level, $E_0$, is then the outer MBSs. However, in the trivial phase the zero-energy crossing in the QD case is accompanied by an exchange of spins in the occupied state. Such spin exchange does not occur in the other cases, leading to a fundamental difference in the spin properties of the QD and PB cases, even if they both host discrete low-energy states below the quasi-continuum.

We finally analyze the size of the regime where trivial zero-energy QD states are observed. In Fig.~\ref{Fig2}(e,f)  we plot $E_0$ as a function of $\Gamma$ and $B$ for the cases in Fig.~\ref{Fig2}(a,c), respectively. From the low-energy spectrum, we identify the topological phase transition (green line) and the beginning of the zero-energy state QD regime (red line).
The QD regime forms a triangular region which is clearly enlarged with $\Gamma$. 
Remarkably, Fig.~\ref{Fig2}(e) shows that even wide SCs can host a QD regime with trivial zero-energy states for strong enough couplings (white dotted line marks $\Gamma =0.7$ from the ideal case in Fig.~\ref{Fig2}(a)). We thus conclude that SNS junctions readily form natural QDs hosting trivial zero-energy states in the strong coupling regime.

\begin{figure}[!t]
	\centering
	\includegraphics[width=.48\textwidth]{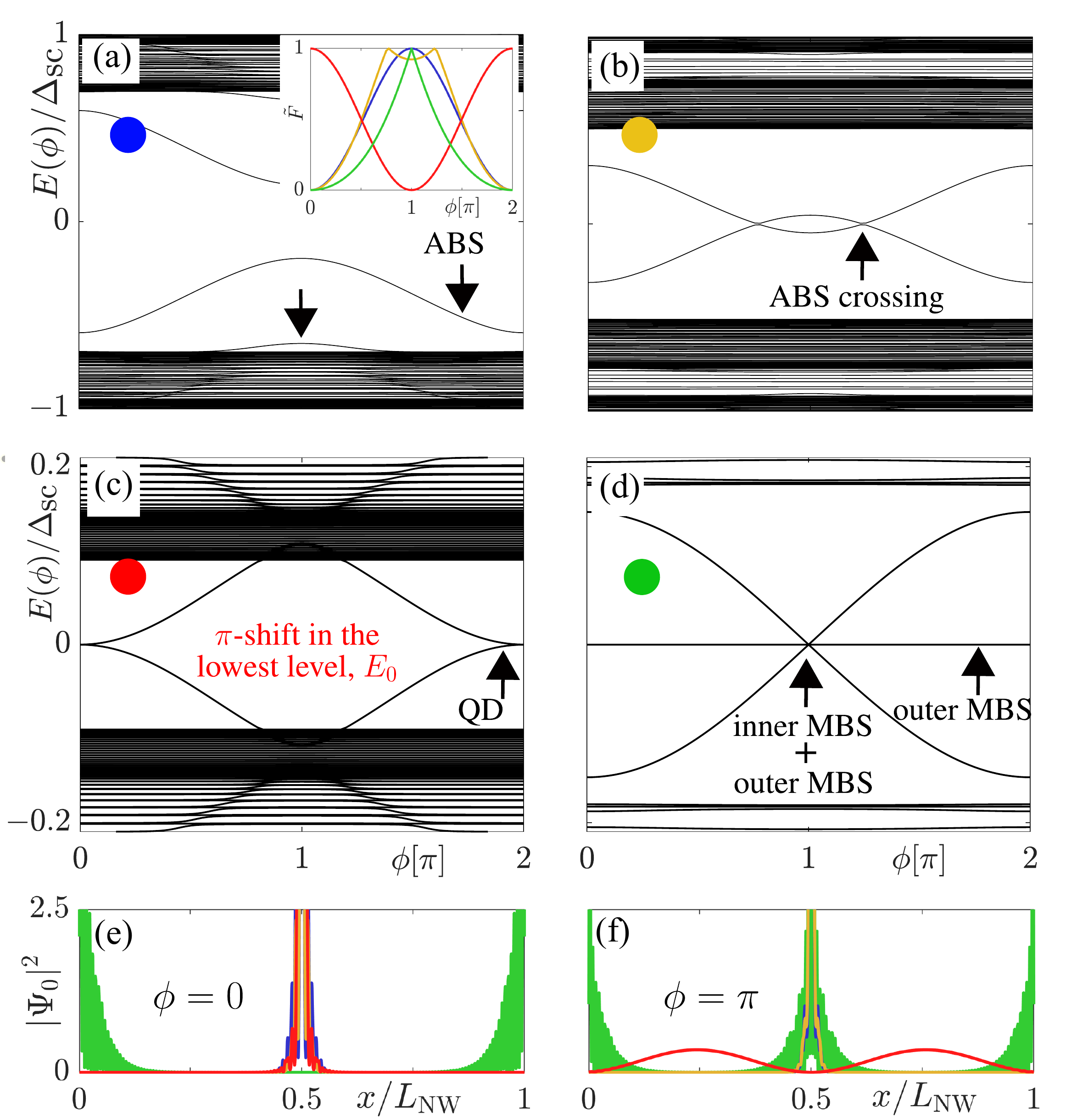}
	\caption{(a-d) Phase-dependent low energy spectrum in the QD case ($L_y=21a$), obtained at the color-marked $B$ values in Fig.~\ref{Fig2}(c). Inset in (a): scaled free energy $\tilde{F}=\left(F-F_{\rm min}\right)/\left(F_{\rm max}-F_{\rm min}\right)$ for (a-d), with $F_{\rm min/max}$ the minimum/maximum of $F$ in each case.  Probability density of the lowest state, $|\Psi_0|^2\cdot 10^3$, in (a-d) at $\phi=0$ (e) and $\phi=\pi$ (f). }
	\label{Fig3}
\end{figure}
%


\emph{Phase-dependence}.--- 
Next we allow for a finite phase $\phi$ across the SNS junction. In particular, we study the phase-dependent energy spectrum for the QD case in Fig.~\ref{Fig2}(c) at the $B$-values identified by the colored bars. At very low $B$ (blue) we find ABSs detached from the quasi-continuum and exhibiting the usual cosine behavior~\cite{Cayao17b,cayao2018andreev}, see Fig.~\ref{Fig3}(a). These lowest energy states are localized at the junction for both $\phi=0,\pi$, see blue line in Figs.~\ref{Fig3}(e,f).
On the other hand, in the topological phase at very large $B$ (green) four MBSs appear in the system: two dispersionless outer MBSs and at $\phi=\pi$ also two MBSs located in the junction (inner MBSs), see Fig.~\ref{Fig3}(d) for the energy spectrum and Figs.~\ref{Fig3}(e,f) for the wave function probabilities.
In both the low $B$ trivial and high $B$ topological regimes, the lowest level reaches maximum negative energy at $\phi=0$. The SNS junction is therefore in the $0$-state because the free energy, $F= \sum_{n<0}E_n$, is minimized at $\phi=0$, see blue and green lines in the inset of Fig.~\ref{Fig3}(a).
 
It is at intermediate $B$ in the trivial phase that dramatic changes takes place. First, the ABSs move towards zero energy with increasing $B$ and start to cross, see Fig.~\ref{Fig3}(b). As a consequence, the free energy, plotted in gold in the inset in Fig.~\ref{Fig3}(a), has a global minimum at $\phi=0$ and a local minimum at $\phi=\pi$. The junction is thus in a $0'$-state \cite{vecino2003josephson}. 
Further increasing $B$ we find that the global and local minima interchanges, eventually reaching the situation in Fig.~\ref{Fig3}(c). Here  the zero-energy crossing is at $\phi = 0$, implying that a full $\pi$-shift has occurred in the low-energy spectrum. As a consequence, this junction is in a $\pi$-state, since the minimum of $F$ is now at $\phi=\pi$, see red Fig.~\ref{Fig3}(a) inset. 
At $\phi=0$ the ABSs are localized at the junction, as in all other cases in the trivial phase, while at $\phi=\pi$ the lowest energy state is completely delocalized because of mixing with the quasi-continuum, see red in Fig.~\ref{Fig3}(e,f). 

We find that the $\pi$-state always emerges when the SNS junction hosts a pair of QD states with zero-energy crossings. In essence this is because the QD forces the ABS to be at or close to zero energy for $\phi =0$. We also note that the QD introduces a phase-dependence for the quasi-continuum, unlike in conventional short junctions \cite{Beenakker:92}.
We have also verified that the ideal and PB cases do not exhibit any $\pi$-states, see SM \cite{SM}. 
Thus, the phase-dependent energy spectrum offers a remarkably clear differentiation between topologically trivial zero-energy QD levels and MBSs.

\begin{figure}[!t]
	\centering
	\includegraphics[width=0.49\textwidth]{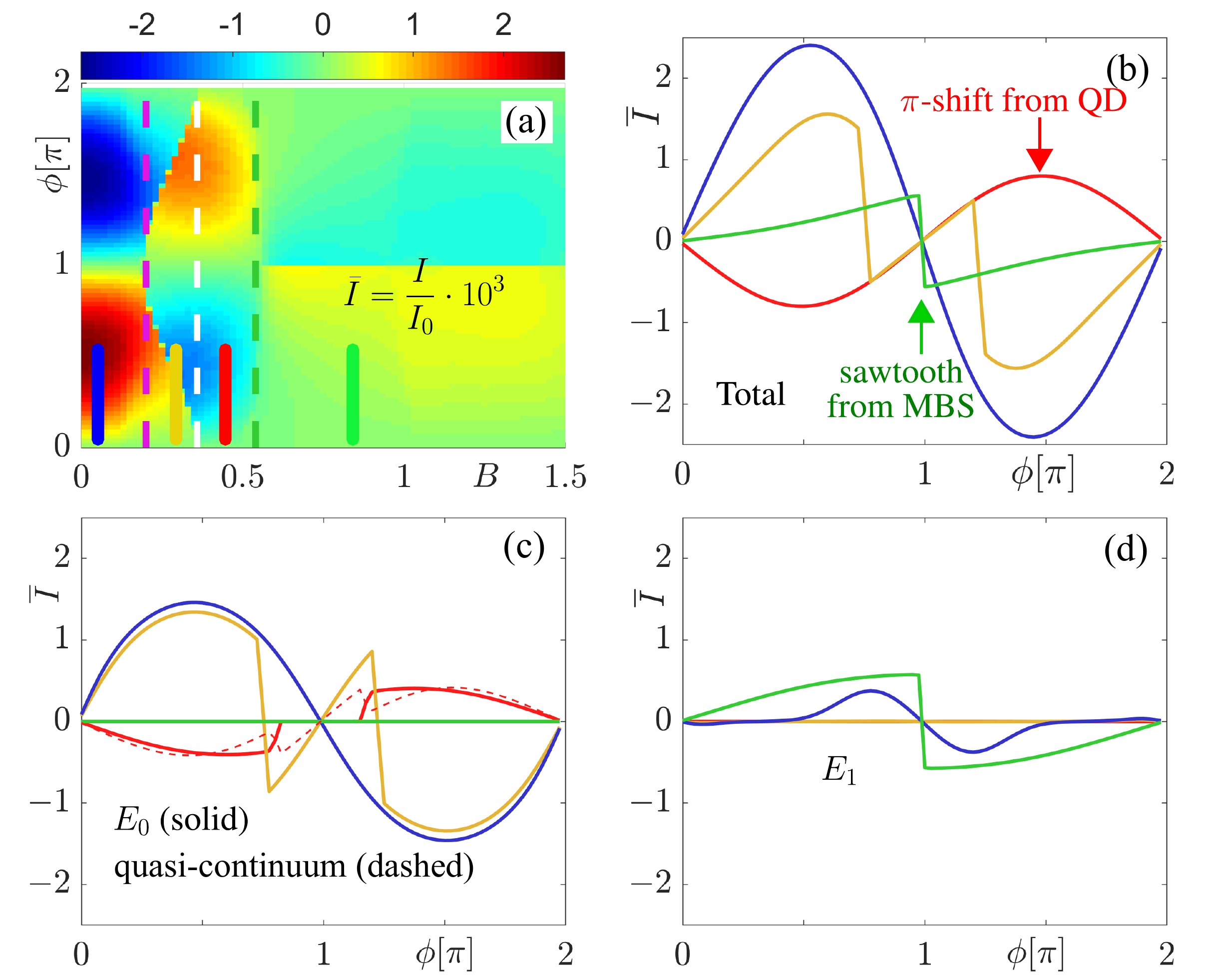}
	\caption[Current]{(a) Colorplot of supercurrent for QD case ($L_y=21a$) as a function of $\phi$ and $B$. Topological phase transition (green dashed line), beginning of ABS crossings in phase-dependent energy spectrum (magenta), and zero-energy crossing at $\phi=0$, i.e.~red arrow in Fig.~\ref{Fig2}(c) (white). Total supercurrent (b), with contributions from $E_0$ (c) and $E_1$ (d) energy levels at the color-marked $B$ values Figs.~\ref{Fig2}(c), repeated in (a).}
	\label{Fig4}
\end{figure}
%
 
\emph{Current-phase relationship}.--- 
To perform a direct detection of the QD trivial zero-energy states we consider the junction supercurrent $I(\phi)$, obtained from $I(\phi)=I_{0}\frac{\partial F}{\partial \phi}$, where $I_0={e}/{\hbar}$. 
Figure~\ref{Fig4}(a) shows a color plot of $\bar{I}=I(\phi)/I_0$ as a function of $\phi$ and $B$ for the QD case in Fig.~\ref{Fig2}(c). For a complete understanding of how the QD levels contribute to $I(\phi)$, we also plot both the total current and the contributions from the lowest ($E_{0}$) and first excited ($E_{1}$) energy levels in Figs.~\ref{Fig4}(b,c,d) for the same $B$ values analyzed in Fig.~\ref{Fig3}.

At low $B$ in the trivial phase $I(\phi)$ displays the usual $\sin(\phi)$-like behavior. This is the $0$-state, where $E_0$ gives the dominating contribution to the supercurrent, albeit $E_1$ also give a small positive contribution, see blue line in Fig.~\ref{Fig4}(b,c,d). 
Beyond the topological phase transition (green dashed line) the situation is also easy to understand. Here $I(\phi)$ has a characteristic sawtooth profile at $\phi = \pi$ due to the special zero energy behavior of the inner MBS at $\phi = \pi$, which has been proposed as a signature of true MBSs in short SNS junctions~\cite{cayao2018andreev,Cayao17b}.

Between the magenta and white lines in Fig.~\ref{Fig4}(a), we find a region with a discontinuous $I(\phi)$, which is caused by the ABS crossings in Fig.~\ref{Fig3}(b). Here, the $E_0$ levels are strongly dispersive with $\phi$ leading to the largest contributions to $I(\phi)$, see gold in Fig.~\ref{Fig4}(c).
Finally, between the dashed white and green lines in Fig.~\ref{Fig4}(a), we find a full sign-reversal for the supercurrent, with the white line corresponding to the red arrow in Fig.~\ref{Fig2}(c) indicating the zero-energy crossing at $\phi=0$. 
This $\pi$-shifted supercurrent arises from the special behavior of the low-energy spectrum: the lowest ABSs exhibit maximum energy at $\phi=\pi$, see Fig.~\ref{Fig3}(c), instead of a minimum as is the case for conventional junctions \cite{Beenakker:92}. Thus the $E_{0}$ level contributes strongly to the $\pi$-shifted supercurrent, as also seen in red in Fig.~\ref{Fig4}(c). 
Due to  the presence of the QD levels, the quasi-continuum also gives a $\pi$-shifted contribution to $I(\phi)$. 
For the ideal and PB junctions, the ABS energy spectrum only exhibits $0$,  $0'$, $\pi'$-states, but never the $\pi$-state and thus we never see a $\pi$-shifted supercurrent. Some signatures of the QD and PB junctions can also be captured by the critical current but not as clear as the $\pi$-shift, see SM \cite{SM}.

For SNS junctions with trivial zero-energy crossings we always find a $\pi$-shifted supercurrent, independent on any zero-energy pinning after the crossing. These zero-energy levels, appearing in the QD regime, are however somewhat sensitive to SOC~\cite{droste2012josephson}, with very large SOC inducing level repulsion, which gaps the spectrum and thus destroys the supercurrent $\pi$-shift, see SM \cite{SM}. 
Interestingly, QD levels in clearly non-topological Josephson junctions have previously been shown to change the state of the junction from $0$ to $\pi$ with increasing magnetic field and also associated with a spin exchange \cite{RevModPhys.77.935,Chang2013Tunneling, Yokoyama2013Josephson, lee2014spin, szombati2016josephson, vecino2003josephson,Zuo2017Supercurrent,Lee2017Scaling,Saldana2019charge}, fully consistent with our findings.


In conclusion, we demonstrate the emergence of zero-energy states in the trivial phase of short SNS NW junctions, due to strong NW-SC coupling causing a QD formation in the NW and tunable by the SC width.  Most significantly, these zero-energy states produce a $\pi$-shift in the phase-biased supercurrent, making them easily distinguishable from MBSs appearing in the topological phase.

\vspace{8pt}
We thank C.~Reeg and C.~Schrade for useful discussions and M.~Mashkoori for helpful comments on the manuscript.
We acknowledge financial support from the Swedish Research Council (Vetenskapsr\aa det), the G\"{o}ran Gustafsson Foundation, the Swedish Foundation for Strategic Research (SSF), the Knut and Alice Wallenberg Foundation through the Wallenberg Academy Fellows program and the EU-COST Action CA-16218 Nanocohybri. Simulations were performed on resources provided by the Swedish National Infrastructure for Computing (SNIC) at the Uppsala Multidisciplinary Center for Advanced Computational Science (UPPMAX).

\bibliography{biblio}
\cleardoublepage
\onecolumngrid
\appendix


\section*{Supplementary material}

# Supplemental Material: Supercurrent detection of topologically trivial zero-energy states in nanowire junctions

Oladunjoye A. Awoga, Jorge Cayao,[*] and Annica M. Black-Schaffer

*Department of Physics and Astronomy, Uppsala University, Box 516, S-751 20 Uppsala, Sweden*

(Dated: August 29, 2019)

In this supplementary material we provide details to further support the results and conclusions of the main text. We first give detailed demonstrations of the effective chemical potential, $\mu_{\rm eff}$, and the renormalization of the spin-orbit coupling (SOC). Then we show how different QD systems, produced by different SC widths or junction lengths, can have varying energy spectra, but still always result in a $\pi$-shifted supercurrent in the short to intermediate junction regime. However, the $\pi$-shift in the supercurrent is destroyed when SOC is extremely large. Then we turn to the potential barrier (PB) case, and show explicitly that the phase-dependent energy spectrum and supercurrent do not support $\pi$-shift. We also show that the trivial zero or near-zero energy levels leave traces in the critical currents, albeit not as clear as the $\pi$-shifted supercurrent. Further, by changing the model parameters we show that our results are robust even at realistic parameters and under self-consistent calculations. Finally, we show that an 1D effective model with spatially inhomogeneous effective chemical but put-in by hand exhibits the same $\pi$-shift feature in the supercurrent.

## Effective chemical potential

In this section we provide more details in order to understand the origin of $\mu_{\rm eff}$, which is plotted in Fig. 1(b) in the main text. For simplicity we consider only one half of the system, that is, instead of a NW coupled to two SCs we only take one of the SCs since the coupling is the same for both SCs. The Hamiltonian is the same as that presented in the main text except now we have a single SC. We further assume that the SC and the NW are infinitely long, with the NW terminated at the edges of the SC, as schematically illustrated in Fig. S1(a). Then the system is translationally invariant along the NW direction so that we can Fourier transform with $k_x$ being a good quantum number, while in the direction perpendicular to NW, $L_y$ is still finite. This approach grants access to the individual reciprocal space subbands of the NW and SC (even after hybridization), making it easy to extract renormalized parameters. Since $\Delta_{\rm sc}$ is a small parameter, we can extract all renormalized parameters in the normal state.

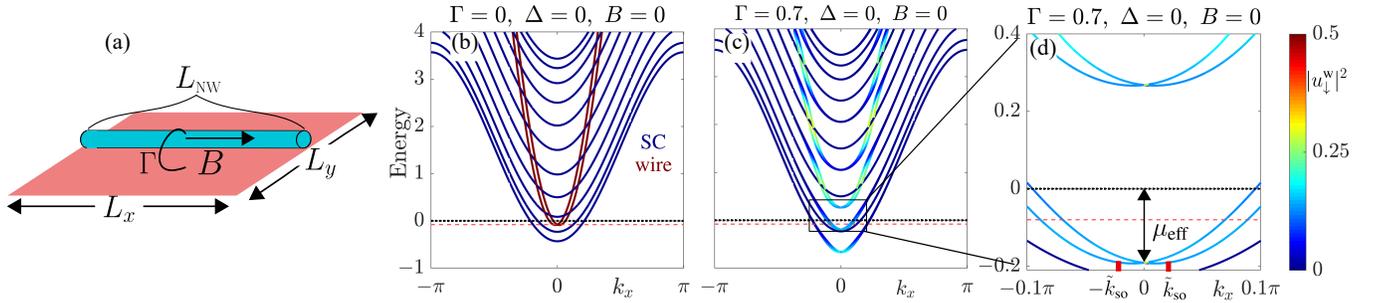

FIG. S1. (a) Schematics of a 1D NW, with length $L_{\rm NW}$ and parallel Zeeman field $B$ coupled to a single 2D SC, of width $L_y$ and length $L_x$, with coupling strength $\Gamma$. Energy subbands for $L_y = 11a$ and $B = 0$ for $\Gamma = 0$ (b) and $\Gamma = 0.7$ (c) with zoom-in at low energies (d). After hybridization the bottom of the subband closest to zero energy (dotted black line) defines the effective chemical potential, $\mu_{\rm eff}$. Here dashed red line marks the bottom of the NW band before hybridization. Minimum of the effective band gives the spin-orbit energy, while the momentum at this point is the renormalized SOC momentum $\tilde{k}_{\rm so}$, indicated by red marker in (d). Subband colors in (b-d) are set by the eigenstate weight in the NW for each subband.

At $\Gamma = 0$, the energy spectrum of the 2D SC consists of a discrete set of doubly (spin) degenerate subbands, due to the finite width, shown for $L_y = 11a$ in Fig. S1(b). The NW, on the other hand, has a pair of non-degenerate bands due to the spin mixing effect of the SOC, except at $k_x = 0$ where the SOC vanishes. This is the situation before any hybridization or renormalization. At finite coupling, $\Gamma \neq 0$, all the parameters of the NW are renormalized as a consequence of NW-SC hybridization [1–4]. In particular, the subbands of the SC and NW are hybridized and shifted in energy with respect to the subbands at $\Gamma = 0$, see Fig. S1(c), where the eigenstate weight in the NW of each band

is coded in color. We find that the band shift directly modifies the native chemical potential of the wire, $\mu_{\rm NW}$, leading to an effective chemical potential $\mu_{\rm eff}$. To a good first approximation $\mu_{\rm eff}$ can be measured at $k_x = 0$ as the distance to the bottom of the subband closest to the Fermi energy (here $E = 0$), since the superconducting gap opens around the Fermi energy, see Fig. S1(c,d). By changing $L_y$, the number of subbands changes, which in turn modifies the spacing between the subbands and thus $\mu_{\rm eff}$. Hence, $\mu_{\rm eff}$ depends heavily on $L_y$. In Fig. 1(b) in the main text we present $\mu_{\rm eff}$ as a function of $L_y$ for $\Gamma = 0.7$.

Ideally, $\mu_{\rm eff}$ should be as close as possible to the Fermi energy in order to easily be able to tune the system into the topological phase [5]. In reality it is however difficult to know a priori the exact $L_y$ that gives such a desired $\mu_{\rm eff}$. Instead we likely obtain any of the three cases discussed in the main text, namely, the ideal, potential barrier (PB), and quantum dot (QD) cases, when considering a short SNS junction with different SC widths. To further explore these three cases, we fix $L_y$ to the values that gave each of the three cases in the main text and plot in Fig. S2 the behavior of $\mu_{\rm eff}$ as a function of coupling strength. Specifically, we use $L_y = 41a$ (triangle) giving the ideal junction in the main text, $L_y = 11a$ (cross) giving the PB junction, and $L_y = 21a$ (dot) giving the QD junction for $\Gamma = 0.7$ used in the main text. As seen, the width of the SC does not at all play a role in the weak coupling regime, but there all systems give a negative $\mu_{\rm eff}$. In the strong coupling, however, $\mu_{\rm eff}$ is very sensitive to $L_y$, thus creating the three different cases. The magnitude of $\mu_{\rm eff}$ also increases with $\Gamma$, thus increasing the possibility of a QD spontaneously forming in the SNS junction. This is in agreement with the results in Fig. 2(e) in the main text. Note that in the weak coupling regime $\mu_{\rm eff}$ is the bottom of the NW subband.

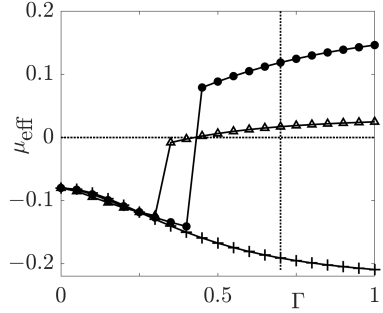

FIG. S2. Effective chemical potential $\mu_{\rm eff}$ as a function of $\Gamma$, for the SC widths $L_y = 41a$ (triangle), $L_y = 11a$ (cross) and $L_y = 21a$ (dot), giving the ideal, PB and QD cases for $\Gamma = 0.7$ (vertical line) used in the main text.

**Effective spin-orbit coupling**

In the main text we give the condition for the existence of a QD solution as $\tilde{\alpha} < B < B_c$ and quoted $\tilde{\alpha} \approx 0.5\alpha_{\rm NW}$ for our choice of parameters, with $\tilde{\alpha}$ being the renormalized SOC strength. Here we demonstrate how $\tilde{\alpha}$ is numerically extracted and show that the SOC in the NW is generally reduced in the NW+SC system.

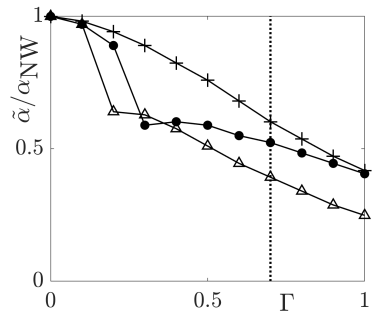

FIG. S3. Renormalized SOC strength $\tilde{\alpha}$ as a function of $\Gamma$ for the SC widths $L_y = 41a$ (triangle), $L_y = 11a$ (cross) and $L_y = 21a$ (dot), giving the ideal, PB, and QD cases at $\Gamma = 0.7$ (vertical line) in the main text. The QD case at $\Gamma = 0.7$ gives $\tilde{\alpha} \approx 0.5\alpha_{\rm NW}$.

To extract $\tilde{\alpha}$, let us first consider the non-hybridized $\Gamma = 0$ case, when $\tilde{\alpha} = \alpha_{\text{NW}}$. The NW band then has $E_{\text{NW}}$ as its minimum at $k_{\text{so}}$. Therefore, $\frac{dE_{\text{NW}}}{dk_x}|_{k_{\text{so}}} = 0$. By solving the resulting equation we obtain $\frac{\alpha_{\text{NW}}}{t_{\text{NW}}} = \tan(k_{\text{so}}a)$. Since $k_{\text{so}}a$ is small we can write $\frac{\alpha_{\text{NW}}}{t_{\text{NW}}} \approx k_{\text{so}}a$. At finite $\Gamma \neq 0$, the minimum of the effective NW subband is instead at $\tilde{k}_{\text{so}}$, see red marker in Fig. S1(d). Since the effective NW band is similar to the NW band at low momentum before the coupling, the expression for the SOC in the effective band is the same as before but now with renormalized parameters such that $\frac{\tilde{\alpha}}{t_{\text{NW}}} \approx \tilde{k}_{\text{so}}a$, which directly give us $\tilde{\alpha}$.

In Fig. S3, we show $\tilde{\alpha}$ for three different $L_y$ as a function of $\Gamma$. As $\Gamma$ increase, $\tilde{\alpha}$ is very clearly reduced. There is also a smaller reduction in $\tilde{\alpha}$ when increasing $L_y$. This means SOC is weakened when coupling the NW to a SC. Note also that since the junction in our system is very short, the SOC at the junction itself must be renormalized in almost the same manner as in the superconducting parts of the wire. Interestingly, this simple analysis provided here gives the same qualitative result as a more elaborate recent work [6].

### Variation of quantum dot levels with superconductor width

In the main text we show that the value of the effective chemical potential $\mu_{\text{eff}}$ is crucial for the emergence of zero-energy QD levels. When $\mu_{\text{eff}}$ is appreciably positive, a QD is induced at the junction and the QD levels can even cross zero energy in the topologically trivial phase. However, the QD level depends $L_y$ since $\mu_{\text{eff}}$ is highly tunable with $L_y$. In Fig. S4(a-d) we show the energy spectrum of the first four $L_y$ values that results in the peak positive values in $\mu_{\text{eff}}$ in Fig. 1(b) in the main text, as representative QD systems.

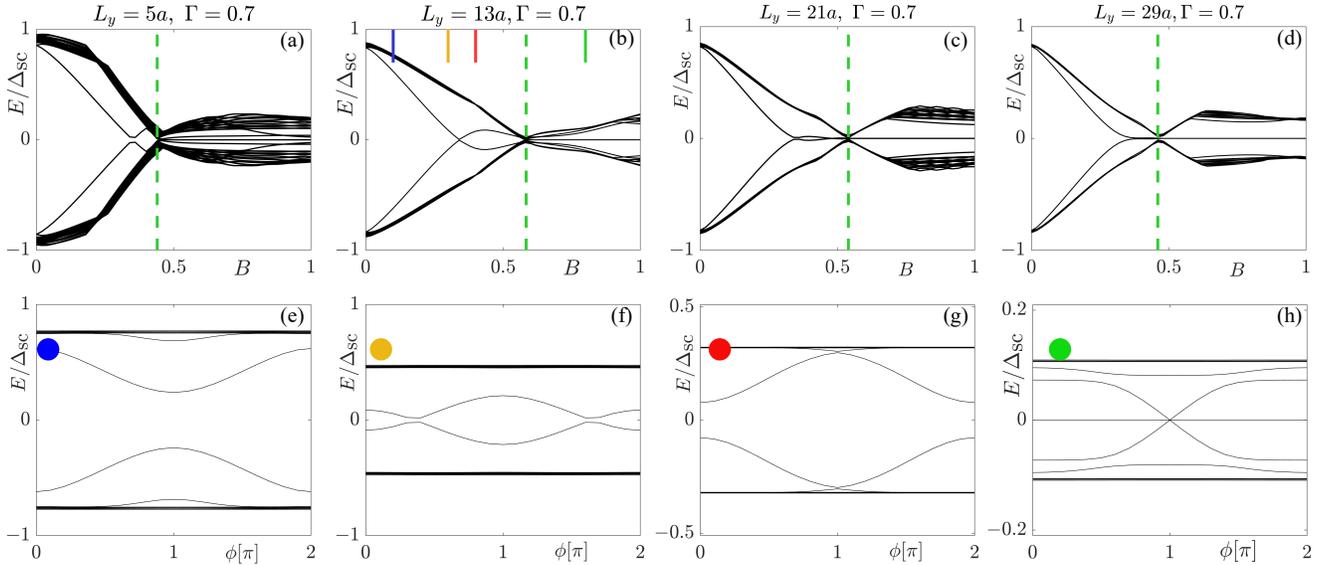

FIG. S4. (a-d) Zeeman field-dependent energy spectrum at $\phi = 0$ (only lowest 30 levels) for the first four $L_y$ values that results in a $\mu_{\text{eff}}$ peak (results in (c) are the same Fig. 2(c) in the main text). Vertical green lines mark the topological phase transition. The QD levels can be pinned around zero as in (c,d), but can also oscillate around zero as in (a,b). (e-h) Phase-dependent spectrum taken at the $B$ values marked by color bars in (b), showing a $\pi$-shift in the lowest level (g).

Despite varying behavior of the QD levels when changing $L_y$, we always find a $\pi$-shift in the phase-dependent spectrum, $E(\phi)$, and thus in the supercurrent $I(\phi)$. This can be understood when considering that the ABS spectrum results in states close to zero energy at $\phi = 0$ but close to the band gap edge at $\phi = \pi$. The spin exchange taking place at the first zero-energy crossing further supports the existence of the $\pi$-shift as it means the ground state of the junction is changed, consistent with results from other non-topological QD junctions [7–12]. To explicitly demonstrate the $\pi$-shift we show in Fig. S4(e-h) the phase-dependent spectrum $E(\phi)$ for the case of $L_y = 13a$, taken at the color markings in Fig. S4(b). Although the QD levels are not at all pinned to zero energy at $\phi = 0$ (in contrast to the $L_y = 21a$ case in the main text), there is still a $\pi$-shift in the lowest level in the trivial phase, see Fig. S4(g). This $\pi$-shift in the lowest level gives rise to a $\pi$-shift in the supercurrent, as demonstrated in Fig. 4 in the main text.

**Increased junction length**

In the main text we used a short junction with length of $2a$, to be compared to the the superconducting coherence length $\xi \approx 13a$ for our choice of parameters. In Figs. S5(a,b) we display the spectrum of the $L_y = 21a$ QD case for two longer junctions. Interestingly, zero-energy QD levels persist even in very long junctions, albeit they do not stay pinned to zero energy. We find, due to the evolution of the spectrum with $\phi$ and coinciding with the exchange of the spin state at every zero-energy crossing of the QD levels, that the state of the junction fluctuates between $\pi$ and 0, with $0'$ and $\pi'$-states in-between. This can numerically be quantified by finding the phase $\phi_{\min}$ at which the $E_0$ level has its minimum, i.e. $E_0(\phi_{\min})$ is the minimum of $E_0(\phi)$, which we plot in Figs. S5(c,d). As seen, there are no direct 0 to $\pi$ transitions, rather the $0 - \pi$ transitions pass through intermediate $0'$ and $\pi'$ states, giving rise to the slightly slanted lines. Still the $\pi$-state is the dominating state across the QD regime.

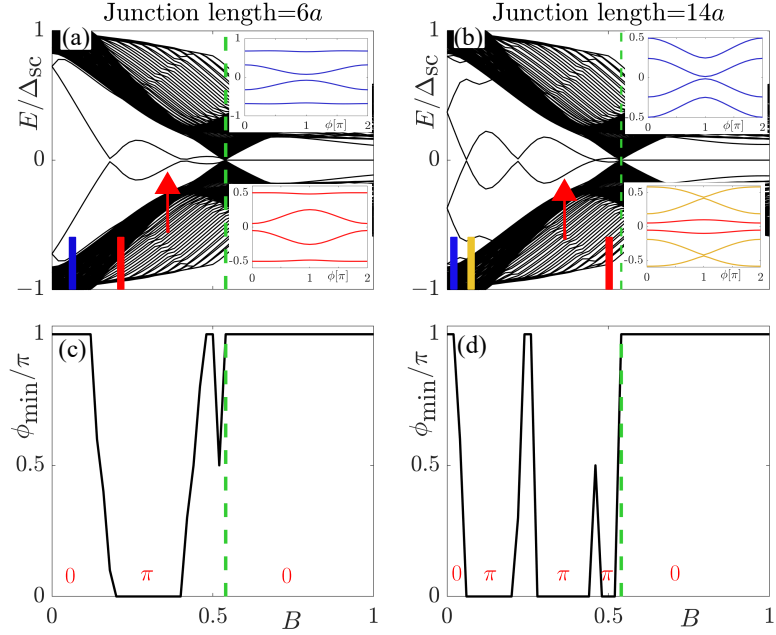

FIG. S5. Zeeman field-dependent energy spectrum at $\phi = 0$ (only lowest 300 levels) for the $L_y = 21a$ QD case for different long junction lengths, $6a \approx \frac{\xi}{2}$ (a) and $14a \approx \xi$ (b), with the phase at the minimum of the lowest level $E_0(\phi)$, $\phi_{\min}$, (c,d). Vertical green lines mark the topological phase transition, while red arrows indicate the original starting point of QD levels for the $2a$ short junction, i.e. same red arrow as in Fig. 2(c) in the main text. Insets show phase dependent spectrum of the lowest levels at the color markings in (a,b).

Furthermore we find that the phase-dependent spectra, insets in Figs. S5(a,b), exhibit $\pi$-shifts in the lowest state but with flatter dispersions compared with the short junction, Fig. 3 in the main text. Very long junctions even allow for multiple levels, with counter-dispersing phase dependences on the energy, see in insets of Fig. S5(b), which weaken the $\pi$-shift in the supercurrent. Thus the $\pi$-shift is most reliable as a tool to distinguish QD zero-energy levels in short SNS junctions.

**Effects of spin-orbit coupling**

We here further clarify the effects of SOC on the QD zero-energy levels. We consider $L_y = 21a$ which gives QD levels as shown in Fig. 2(c) in the main text. The effect of SOC is shown in Fig. S6. First, we set $\alpha_{\text{NW}} = 0$. In this case there is no opening of the topological gap since there is no SOC, see Fig. S6(a). However, the QD levels still exist before the putative gap closure. The $\pi$-shift in the supercurrent also remains (not shown), beginning at the $B$-value for the zero-crossing of the QD levels. Note that the condition for QD levels leading to $\pi$-shift in the supercurrent $\tilde{\alpha} < B < B_c$ clearly holds in this case.

Next we set $\alpha_{\text{NW}} = 0.1 t_{\text{NW}}$, i.e. twice the value in the main text. We here find that the QD levels do not reach zero energy, see Fig. S6(b). Instead, the large $\alpha_{\text{NW}}$ value introduces a large hybridization between the QD levels

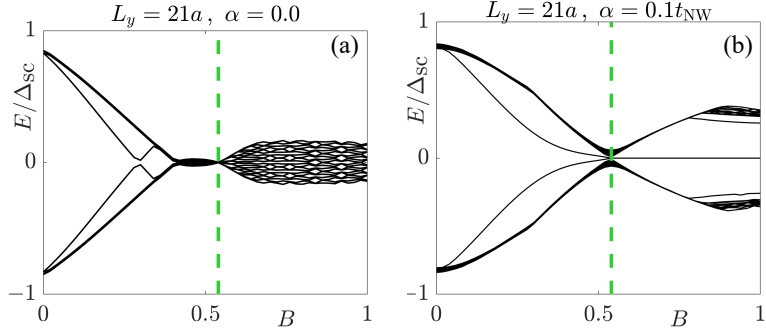

FIG. S6. Zeeman field-dependent energy spectrum (only lowest 30 levels) for $\alpha_{\rm NW} = 0$ (a) and $\alpha_{\rm NW} = 0.1 t_{\rm NW}$ (b) for the $L_y = 21a$ QD case. In (a) there is no topological phase transition, while in (b) the QD levels do not reach zero energy due to a too large $\alpha_{\rm NW}$, results to be compared with Fig. 2(c) in the main text, where $\alpha_{\rm NW} = 0.05 t_{\rm NW}$.

leading to anti-crossings. In this case, the $\pi$-shift is also not observed in the system, but we only find a discontinuous supercurrent. For this very large $\alpha_{\rm NW}$ the renormalized value of SOC $\tilde{\alpha}$ in the NW is also large and violates the condition for zero energy QD levels, $\tilde{\alpha} < B < B_c$. This behavior of non-zero-energy QD states at high SOC is also consistent with earlier reports of QD levels in non-topological SOC systems [13].

### Potential barrier case

As stated in the main text we do not find a $\pi$-shift in the phase-dependent energy spectrum or supercurrent for the potential barrier (PB) case. For completeness we present in Fig. S7(a-d) the phase-dependent spectrum at four different $B$ values for the PB case occurring at $L_y = 11a$ (see Fig. 2(b) in main text for the Zeeman-field dependent spectrum at $\phi = 0$, with the $B$ values chosen by the corresponding color indicated values in Fig. 2(c)). As seen in Fig. S7(b,c), the lowest states never cross zero energy close to $\phi = 0$ so there cannot be a $\pi$-shift in the lowest levels.

As expected from the features of the phase-dependent spectrum, the supercurrent does not either have have a $\pi$-shift, see Fig. S7(e,f). The individual contributions to the current from the $E_0$ and $E_1$ levels are for completeness also shown in Fig. S7(g,h). Note that close to $\phi = \pi$, for $B$-values between the magenta and green dashed lines in the trivial phase, there is a sign change in the supercurrent in Fig. S7(e). However, this sign change does not occur throughout the whole $\phi$ range, and it therefore does not correspond to a $\pi$-shifted current. Instead, this sign change only occurs in a limited range of $\phi$, which only renders the supercurrent discontinuous, a consequence of the ABS crossings in Fig. S7(b,c).

### Critical currents

Beyond the current-phase relationship of the current, another experimentally relevant quantity is the critical current, $I_c$, which represent the maximum supercurrent that flows across the junction. Critical currents can provide additional insight to current-phase relationships that we present in the main text. Technically we extract the $I_c$ by maximizing the supercurrent $I(\phi)$ with respect to the superconducting phase difference $\phi$. The supercurrents for the PB and QD cases are shown in Fig. S8. In an ideal situation the critical current decreases with increasing $B$ and traces out the topological gap closure [14]. Remarkably, in the PB and QD cases the low-lying states in the trivial phase also modifies $I_c$. In the PB case the critical current is low in much of the trivial phase, see Fig. S8(a). This can be understood from having close lying levels, see Fig. S7(b,c), with counter-dispersive phase-dependent energy spectrum, thereby contributing destructively to the total supercurrent, hence, giving a low critical current. Note how the junction transitions between intermediate states but never reaches the full $\pi$-state such that only a discontinuous current-phase relationship is found in the PB case.

For the QD case we instead find a bump in $I_c$ within the QD region, between the black and green lines in Fig. S8(b). This bump is a telltale of the zero-energy QD levels carrying most of the supercurrrent. The dip at the topological transition comes from the energy gap closing. Note also the transition between intermediate states before reaching the $\pi$-state. In general, we find that the critical current shows a kink or plateau whenever there is a transition between different states of the junction.

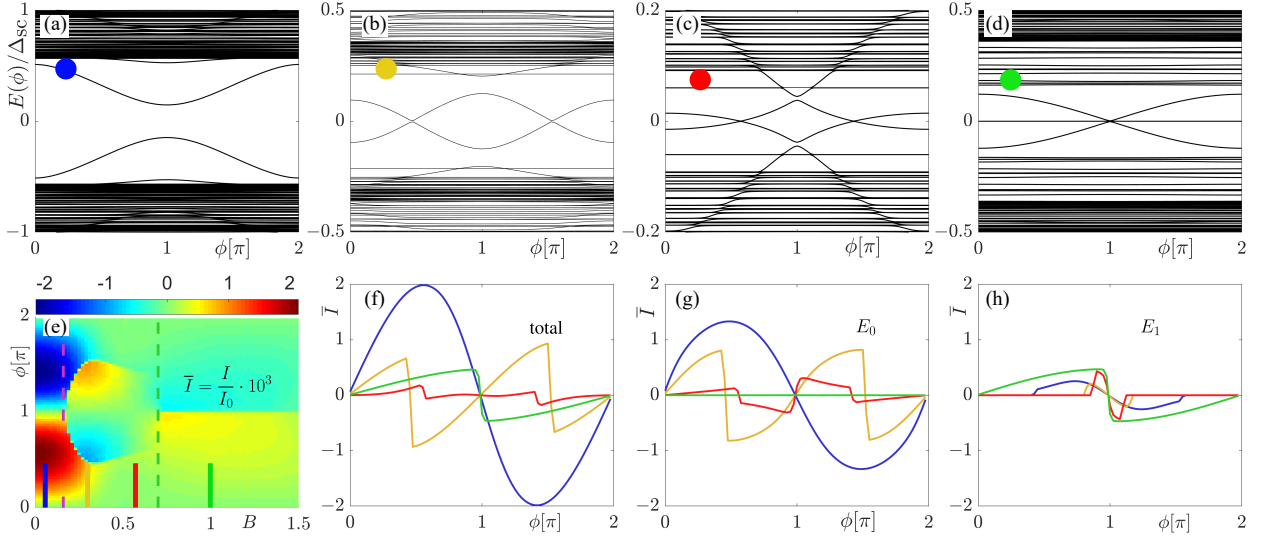

FIG. S7. (Top row) Phase-dependent energy spectrum for the $L_y = 11a$ PB junction obtained at the color-marked $B$ values in panel (e). Compare with Fig. 3 and 4 for the QD case in the main text. (Bottom row) Colorplot of supercurrent as a function of $\phi$ and $B$ (e), total supercurrent (f), with contributions from the $E_0$ (g) and $E_1$ (h) energy levels taken at $B$ values marked with the corresponding color bars in (e). Note that there is no $\pi$-shift in the supercurrent.

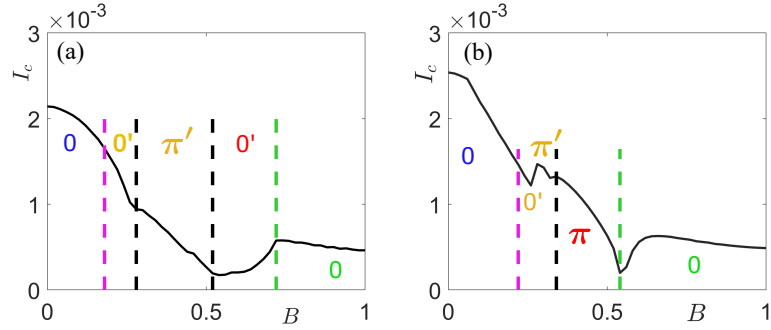

FIG. S8. Zeeman field-dependent critical current $I_c$ for $L_y = 11a$ PB (a) and $L_y = 21a$ QD (b) cases. Magenta and green dashed lines mark the beginning of the ABS zero-crossing (at $\phi = \pi$) and topological phase transition, respectively. Black dashed lines mark transitions to intermediate states $0'$ and $\pi'$.

## TOWARDS MORE REALISTIC PARAMETERS

In the main text we used an somewhat overestimated value of the superconducting order parameter, $\Delta_{\rm sc}$, in the parent SC and as a consequence a large coupling strength $\Gamma$ is needed to reach the strong coupling regime. We here reduce the order parameter by a factor of 5, such that $\Delta_{\rm sc} = 0.02$, in order to be close to realistic values. For this value of $\Delta_{\rm sc}$, strong coupling is achieved already around $\Gamma = 0.3$, see Fig. S9(b), when also using the reduced chemical potentials, $\mu_{\rm sc} = 0.2$ and $\mu_{\rm NW} = 0$ to keep close to realistic values. As a consequence of the longer superconducting coherence length in the SC we also increase the system size and junction length. Since $\mu_{\rm eff}$ depends on the Fermi momentum of the SC, we find that $\mu_{\rm eff}$ changes with the chemical potential of the SC but the overall behavior does not change, compare Fig. S9(a) with Fig. 1(b) in the main text.

Similar to the main text we consider three SC widths yielding different $\mu_{\rm eff}$, see markers in Fig. S9(a), representing the ideal, PB, and QD junction behaviors. We find the same behavior in the low energy spectrum as in Fig. 2 in the main text: the topological phase transition point changes with $L_y$ and there are trivial zero-energy states for the QD case and also near zero-energy states in the PB case, see Fig. S11(b,c).

The phase dependent energy spectrum for the QD case ($L_y = 49a$) is shown in Fig. S11. We find that the $\pi$-shift persists in the QD regime, see Fig. S11(c). This $\pi$-shift in the lowest level gives a $\pi$-shift in the supercurrent (not shown). Furthermore, we find only $0'$, $\pi'$ states but no $\pi$ state in the PB case (not shown), despite the energy levels

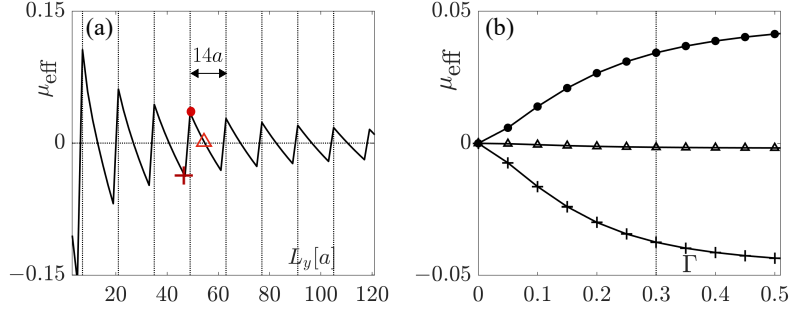

FIG. S9. Effective chemical potential profile deep into the SC parts of the NW as a function of $L_y$ for fixed $\Gamma = 0.3$ (a) and as a function of coupling strength $\Gamma$ for $L_y = 55a$ (triangle), $L_y = 47a$ (cross) and $L_y = 49a$ (dot), representative of the ideal, PB, and QD cases, respectively. Here $\Delta_{\rm sc} = 0.02 t_{\rm sc}$, $\mu_{\rm s} = 0.2 t_{\rm sc}$, $\mu_{\rm NW} = 0$, where $t_{\rm sc} = 25{\rm meV}$ and junction length $= 4a$.

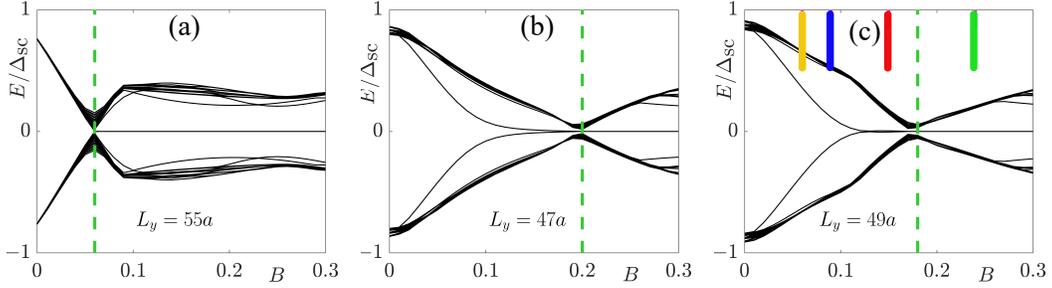

FIG. S10. Zeeman field-dependent energy spectrum (only lowest 30 levels) at $\phi = 0$ for ideal ($L_y = 55a$) (a), PB ($L_y = 47a$) (b), and QD ($L_y = 49a$) (c) cases, given by the markers in Fig. S9. Vertical dashed green lines mark the topological phase transition. Here $\Gamma = 0.3 t_{\rm sc}$, $L_x = 775a$, $L_{\rm NW} = 1500a$, with other parameters the same as Fig. S9. Compare with Fig. 2(a-c) in the main text.

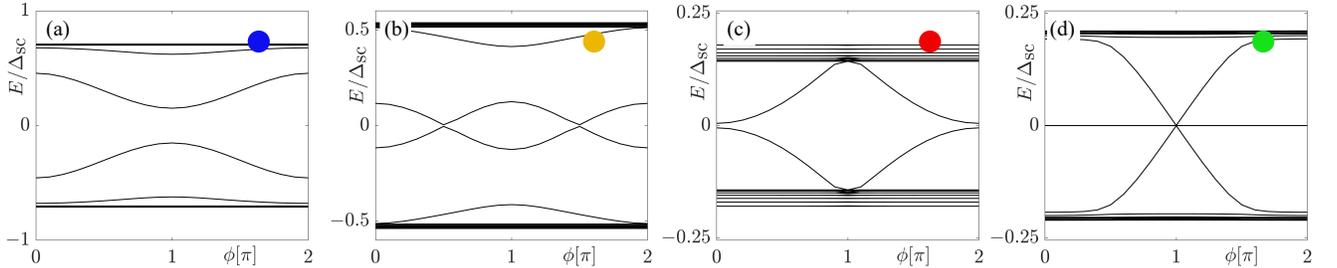

FIG. S11. Phase-dependent energy spectrum (only lowest 30 levels) in the QD case ($L_y = 49a$) obtained at the color-marked $B$ values in Fig. S10(c). Compare with Fig. 3 in the main text.

almost reaching zero energy close to the topological phase transition for the PB case.

## SELF-CONSISTENCY

Hitherto we have taken $\Delta_{\rm sc}$ to be simple constant. It would however be more accurate to calculate $\Delta_{\rm sc}$ self-consistently in the SC through the self-consistency equation $\Delta_{\rm s}(i) = -V_{\rm sc} \langle c_{i\uparrow} c_{i\downarrow} \rangle$, which allows a site dependence for superconducting order parameter, since the NW will also affect the SC [15, 16]. In this calculation, the pair potential $V_{\rm sc}$ is chosen to be constant as it represents a constant tendency for pairing in the SC. To make self-consistent calculations computationally feasible we have to limit ourselves to smaller systems and thus we need to increase $\Delta_{\rm sc}$ such that the superconducting coherence length becomes smaller than the total system size. This is done by using a $V_{\rm sc}$ value that results in large $\Delta_{\rm sc}$.

We find that, qualitatively, self-consistency yields the same result as a constant $\Delta_{\rm sc}$. Except for slightly different

topological phase transition points and QD regimes both calculations match well, as seen in Fig. S12(a). We have also verified that the supercurrents exhibit the $\pi$-shift in the QD regime (not shown). We also study the induced superconducting correlations in the NW. In Fig. S12(b) we plot the spin-singlet pair correlation in the NW normalized with that of the SC: $F_{\downarrow\uparrow} = \langle d_{i\uparrow}d_{i\downarrow}\rangle/|\langle c_{0\uparrow}c_{0\downarrow}\rangle|$, where $\langle c_{0\uparrow}c_{0\downarrow}\rangle$ is calculated deep into the SC where $\Delta_{\text{sc}}$ is constant. As seen, there is a strong suppression of the NW pair correlations in the junction region, with disturbances reaching reasonably far into the S regions. Notably there is also a strong variation with $\phi$. We also study the same NW pair correlations for the non-self-consistent case and find qualitatively the same results, despite the order parameter $\Delta_{\text{sc}}$ being constant in the SC.

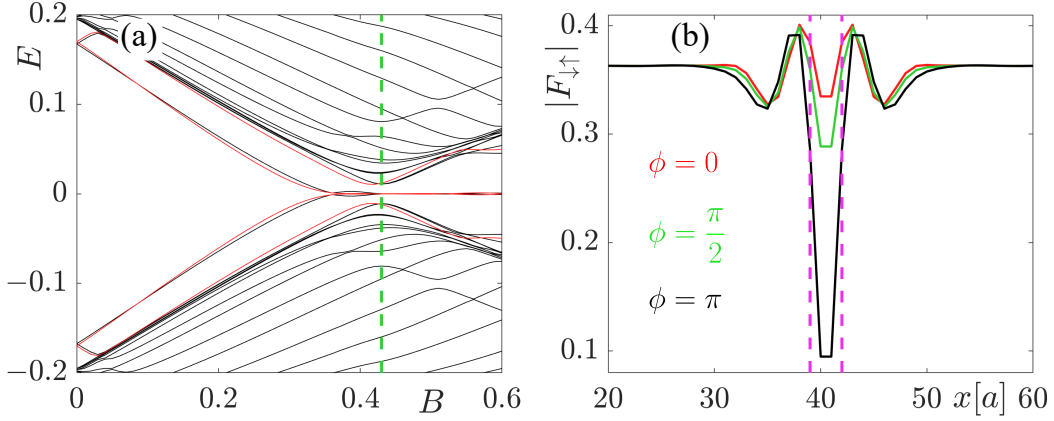

FIG. S12. (a) Zeeman field-dependent energy spectrum at $\phi = 0$ for self-consistent (black) and non-self-consistent (red) calculations for the QD ($L_y = 21a$) case. Vertical dashed green line marks the self-consistent topological phase transition. In the topological phase the zero energy states (MBSs) for the self-consistent (black) calculation are behind the red lines. Magnitude (b) of the induced superconducting pair correlations in the NW at $B = 0$ normalized by the correlation in the parent SC, $F_{\downarrow\uparrow}$. Dashed magenta line mark the junction region. Here $V_{sc} = 0.3$ giving $\Delta_{\text{sc}} = 0.3$, in the bulk SC, $\mu_{\text{s}} = 0.5, \mu_{\text{NW}} = 0.02, \Gamma = 0.75, L_x = 45a$ and $L_{\text{NW}} = 80a$.

## EFFECTIVE 1D MODEL

In the main text and so far in this supplementary material, we have considered a NW coupled to two 2D SCs and solved for the full system. Now, we consider the equivalent setup for a pure 1D junction: a NW with Rashba SOC, parallel magnetic field $B$, and use an assumed induced onsite superconducting order parameter, $\Delta_0$, which we put by hand in the left and right sectors of the NW along with a superconducting phase difference $\phi$ across the junction. We again consider three different cases, ideal, PB, and QD junctions, that correspond to the ones investigated in the NW+SC model. Here we however have to put by hand the values of the chemical potentials in the left/right and central regions of the 1D system that give these three cases. Specifically, the cases are: (1) The chemical potential in the wire is constant, ideal case. (2) The chemical potential at the junction is higher than that of the superconducting parts, PB case. (3) The chemical potential at the junction is lower than that of the superconducting parts, QD case. The Hamiltonian for this pure 1D system is

$$\mathcal{H}_{1D} = \sum_{x=1,\sigma\sigma'}^{L_{\text{NW}}} \left[\varepsilon_{\text{NW}}(x)d^\dagger_{x\sigma}d_{x\sigma} + Bd^\dagger_{x\sigma}\sigma^x_{\sigma\sigma'}d_{x\sigma'} + \Delta(x)d_{x\downarrow}d_{x\uparrow}\right] + \sum_{x=1,\sigma}^{L_{\text{NW}}-1}\left[-t_{\text{NW}}d^\dagger_{x\sigma}d_{x+1,\sigma} + \alpha_{\text{NW}}\left(d_{x\uparrow}d^\dagger_{x+1,\downarrow} - d^\dagger_{x,\downarrow}d_{x+1\uparrow}\right)\right] + \text{H.c.},$$

where $\varepsilon_{\text{NW}}(x) = 2t_{\text{NW}} + \mu_{\text{NW}}(x)$, and $t_{\text{NW}}$ with $\mu_{\text{NW}}$ the nearest neighbor hopping and chemical potential, respectively, while $\alpha_{\text{NW}}$ is the SOC strength and $B$ the Zeeman field. The assumed induced order parameter $\Delta_0$ and the effective chemical potential $\mu_{\text{eff}}$ are further defined as:

$$\{\Delta(x),\ \mu_{\text{NW}}(x)\} = \begin{cases} \{|\Delta_0|e^{i\phi},\ \mu_{\text{eff}}\} & x < L_{\text{s}} \\ \{0,\ 0\} & L_{\text{s}} < x \leq L_{\text{s}} + d\ , \\ \{|\Delta_0|,\ \mu_{\text{eff}}\} & x > L_{\text{s}} + d \end{cases}$$

where $d = 4a$ is the length of the junction, $\phi$ the phase difference across the junction and $L_s$ the length of each of the superconducting parts, assumed to be large. That is, this setup has a superconducting order and a modified chemical potential in the superconducting regions of the NW, modeling what actually happens in the NW+SC system. In particular, we use values of $\mu_{\rm eff}$ extracted from Fig. 1(b) in the main text in order to as closely as possible model strong coupling regime of the NW+SC.

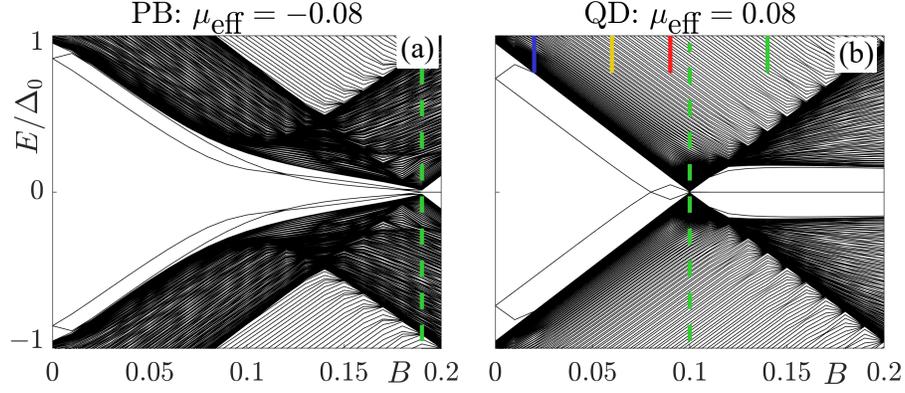

FIG. S13. (a) Zeeman field-dependent energy spectrum at $\phi = 0$ for PB (a) and QD (b) cases for a purely 1D system. Here $\Delta_0 = 0.1$, $\alpha_{\rm NW} = 0.02 t_{\rm NW}$ and $L_{\rm NW} = 2000a$. Vertical dashed green line marks the topological phase transition.

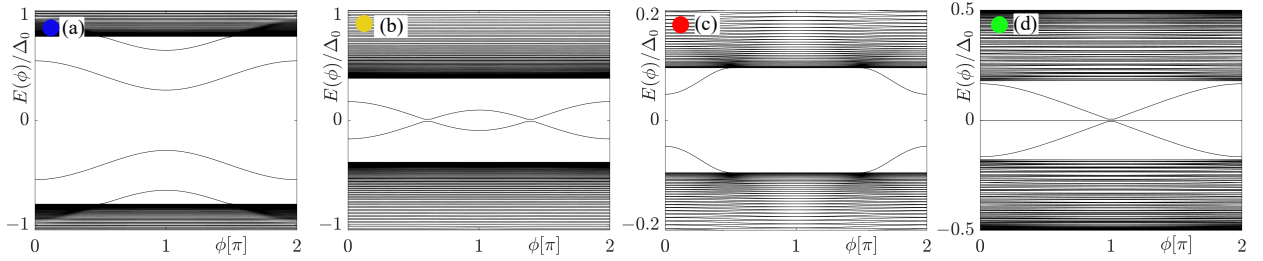

FIG. S14. Phase-dependent energy spectrum for the QD case obtained at the color-marked $B$ values in Fig. S14.

We proceed by numerically diagonalize $\mathcal{H}_{\rm 1D}$ within the Bogoliubov-de Gennes formalism, as in the main text. First, we calculate the evolution of the energy spectrum with magnetic field $B$ for the PB and QD cases, see Fig. S13. As seen, we obtain the same result as for the full NW+SC system, with the trivial phase hosting zero energy states in the QD case. We also find the $\pi$-shift in the phase-dependent energy spectrum for the QD case, see Fig. S14, but not in the PB case (not shown). Thus the existence of the QD zero-energy levels and the associated $\pi$-shifted supercurrent are not due to modeling the full NW+SC system per see, although the QD is produced automatically when considering the full NW+SC, while it has to be put in by hand in the 1D system.

It is here also worth mentioning that we did not find the QD behavior for experimentally realistic values of $\alpha_{\rm NW}$ and $\Delta_0$. We instead had to use a weak $\alpha_{\rm NW} = 0.02 t_{\rm NW}$ to obtain the QD behavior in Fig. S14(b). This is because, as discussed earlier, SOC is significantly weakened in the NW+SC system. Thus, using the original SOC strength in the NW for the 1D effective model is equivalent to a very large native SOC, which lead to level repulsion and gapping out the QD levels, as shown above for the full NW+SC system. This could, perhaps, be the reason why previous work on 1D models did not find the $\pi$-shift in the supercurrent, when we find it to be remarkably robust in the NW+SC system.

---


\end{document}